\documentclass{iaus}
\usepackage{graphics}
  \checkfont{eurm10}
  \iffontfound
    \IfFileExists{upmath.sty}
      {\typeout{^^JFound AMS Euler Roman fonts on the system,
                   using the 'upmath' package.^^J}%
       \usepackage{upmath}}
      {\typeout{^^JFound AMS Euler Roman fonts on the system, but you
                   dont seem to have the}%
       \typeout{'upmath' package installed. iaus.cls can take advantage
                 of these fonts,^^Jif you use 'upmath' package.^^J}%
      }
  \else
  \fi

  \checkfont{msam10}
  \iffontfound
    \IfFileExists{amssymb.sty}
      {\typeout{^^JFound AMS Symbol fonts on the system, using the
                'amssymb' package.^^J}%
       \usepackage{amssymb}%

      }{}
  \fi

  \IfFileExists{amsbsy.sty}
    {\typeout{^^JFound the 'amsbsy' package on the system, using it.^^J}%
     \usepackage{amsbsy}}
    {}

\newsavebox{\astrutbox}
\sbox{\astrutbox}{\rule[-5pt]{0pt}{20pt}}

\title[The Interplay among Black Holes, Stars and ISM in Galactic
       Nuclei]{The mass estimate in narrow-line Seyfert 1 galaxies}

\author[Bian \& Zhao]{W. Bian$^{1,2}$ \& Y. Zhao$^{1}$}

\affiliation{$^{1}$National Astronomical Observatories, Chinese
Academy of Sciences, Beijing 100012, China
\\$^{2}$Department of Physics, Nanjing Normal University, Nanjing 210097, China}

\pubyear{2004} \volume{222} \pagerange{1--8}
\date{?? and in revised form ??}
 \jname{The Interplay among Black Holes, Stars
and ISM \\in Galactic Nuclei} \editors{Th. Storchi Bergmann, L.C.
Ho \& H.R. Schmitt, eds.}
\begin{document}

\maketitle

\begin{abstract}
It is possible that narrow-line Seyfert 1 galaxies (NLS1s) are in
the early stage of active galactic nuclei (AGNs) evolution. It is
important to estimate the mass of supermassive black hole (SBH) in
NLS1s. Here we considered the different kinds of methods to
estimate the SBH masses in NLS1s. The virial mass from the
H$\beta$ linewidth assuming random orbits of broad line regions
(BLRs) is consistent with that from the statured soft X-ray
luminosity, which showed that most of NLS1s are in the
super-Eddington accretion state. The mass from the [O~III]
linewidth is systematically larger than that from above two
methods. It is necessary to measure he bulge stellar dispersion
and/or bulge luminosity in NLS1s.
%\vglue-1.0cm
\end{abstract}

\section{Black Hole Mass Estimate}
There appears a rapid progress on the mass estimate of the
supermassive black hole (SBH) in active galactic nuclei (AGNs),
which boosts our understanding of their central structure and
their evolution (Bian 2004). There is a subclass of AGNs called
narrow-line Seyfert 1 galaxies (NLS1s) (Boller et al. 1996), which
is proposed to be in the early stage of AGNs evolution (Mathure
2000). A popular model of NLS1 is that they contain less massive
black holes with super-Eddington accretion, which leads to the
steep X-ray slope founded in NLS1s (Bian \& Zhao 2004a). To test
this hypothesis we need to estimate the black hole mass of NLS1s.

There are many methods to calculate the black hole masses in AGNs
(Bian \& Zhao 2004b and reference therein): (1)Virial mass derived
from the FWHM(H$\beta$) (or other broad line) and the sizes of
broad line regions (BLRs) from the reverberation mapping technique
or the empirical size-luminosity formula. (2)The relation between
the mass and the bulge stellar velocity dispersion (or [O~III]
linewidth) (M-$\sigma$ relation) or the bulge luminosity
(M-$L_{bulge}$ relation). (3)Soft X-ray variability.  (4)The mass
from spectral energy distribution (SED) fit from the accretion
disk model, such as the method using the soft X-ray hump
luminosity for AGNs with super-Eddington accretion rate. (Wang \&
Netzer 2003).

\section{Data and Results}

Up to now, there are many NLS1s samples (Bian \& Zhao 2004c). We
used the sample of Grupe et al. (2004) for its completeness. We
calculated the masses in AGNs using three different methods:
virial mass from H$\beta$ line ($M(H\beta)$), the mass from
[O~III] line ($M([O~III])$), and the mass from the soft X-ray hump
luminosity ($M(L_{\rm SX})$). In Fig. 1 we plotted the mass from
the X-ray luminosity versus from H$\beta$ line for the sample of
Grupe et al. (2004). From Fig. 1, we found that SBH masses in
NLS1s can be reliably derived from soft X-ray luminosity for their
super-Eddington accretion. The consistency of $M(H\beta)$ and
$M(L_{\rm SX})$ showed that these two methods are available to
estimate the mass in NLS1s, which is consistent that most of NLS1s
are in the super-Eddington accretion state. $M(H\beta)$ and
$M([O~III])$ (from [O~III] line) are not consistent in the sample
of Grupe et al. (2004) and the sample of NLS1s in SDSS (Bian \&
Zhao 2004b). Therefore, compared with $M(H\beta)$ and $M(L_{\rm
SX})$, $M([O~III])$ is not reliable in NLS1s, which is consistent
with the results that NLS1s deviated from M-$\sigma$ relation
defined in non-active AGNs. For more detail, please read Bian \&
Zhao (2004c).
\\
\\
{\small This work has been supported by the NSFC (No. 10273007;
No. 10273011) and NSF from Jiangsu Provincial Education Department
(No. 03KJB160060).}

\begin{figure}
\centerline{\includegraphics{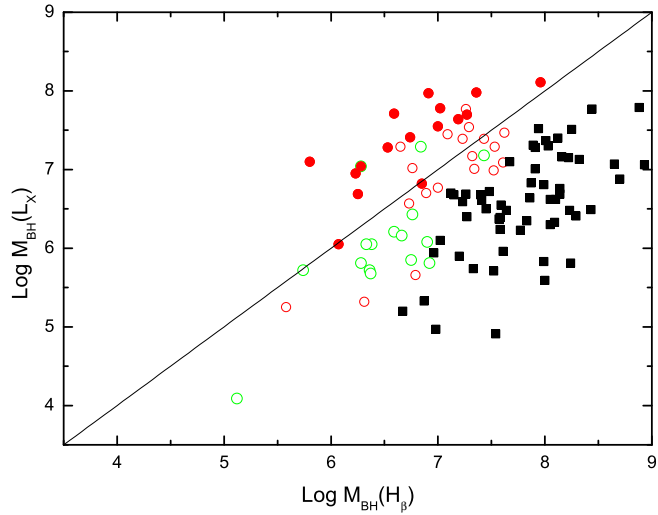}}
%\includegraphics[width=12cm,height=10cm]{f3.eps}
%\centerline{\epsfig{file=f3.eps,width=9cm,angle=0,clip=}}
%\plotfiddle{f3.eps}{7cm}{0}{100}{100}{-150}{10} \vglue-0.8cm

\caption{\small Mass derived from H$\beta$ line versus from the
soft X-ray luminosity for NLS1s and BLS1s in the sample of Grupe
et al. (2004). NLS1s are shown as circles. BLS1s are shown as
solid squares. The green open circles denote NLS1s with $L_{\rm
bol}/L_{\rm Edd}<1$. The red solid circles denote NLS1s which are
satisfied with the Eddington limit relation defined by Wang
(2004). For more detail, please reference Bian \& Zhao (2004c).}

%\vglue-0.5cm
\end{figure}

\end{document}